\documentstyle[aps,preprint]{revtex}
\tightenlines
\def\[{\left\lbrack}
\def\]{\right\rbrack}

\def\({\left(}
\def\){\right)}
\def\ih{\'\i}
\newcommand{\be}{\begin{equation}}
\newcommand{\ee}{\end{equation}}
\newcommand{\ea}{\end{eqnarray}}
\newcommand{\ba}{\begin{eqnarray}}

\begin{document}

\title{Gauging the SU(2) Skyrme model}

\author{J.Ananias Neto,  C.Neves and W.Oliveira \thanks{e-mails:jorge@fisica.ufjf.br,
wilson@fisica.ufjf.br, cneves@fisica.ufjf.br}}

\address{Departamento de F\ih sica, ICE, \\ Universidade Federal de Juiz de Fora, 36036-330, \\ Juiz de Fora, MG, Brazil }

\pagestyle{myheadings}
\markright{J.A.Neto, C.Neves and W.Oliveira, `Gauging the SU(2) Skyrme model'}
\date{\today}

\maketitle
\draft

\begin{abstract}
\noindent In this paper the SU(2) Skyrme model will be reformulated as a gauge theory and the hidden symmetry will be investigated and explored in the energy spectrum computation. To this end we purpose a new constraint conversion scheme, based on the symplectic framework with the introduction of Wess-Zumino (WZ) terms in an unambiguous way. It is a positive feature not present on the BFFT constraint conversion. The Dirac's procedure for the first-class constraints is employed to quantize this gauge invariant nonlinear system and the energy spectrum is computed. The finding out shows the power of the symplectic gauge-invariant formalism when compared with another constraint conversion procedures present on the literature. 
\end{abstract}
   
\maketitle

\newpage

\section{Introduction}

We unveil the hidden symmetry of the SU(2) Skyrme model\cite{Skyrme} lying on the original phase-space. It is a new conception not yet investigated. This hidden symmetry will be investigated using the symplectic gauge-invariant formalism. This new technique, developed by us in this paper, reformulates noninvariant models as gauge invariant theories. 

The SU(2) Skyrme model is an effective theory that describes the weakly interacting mesons in the chiral limit resulting from the more fundamental theory for strong interactions(QCD) in the limit when the number of colors $N_c$ is taken very large. The collective semi-classical approach \cite{Adkins,ANW} leads to the isospin quantum corrections to the baryons properties. This process reduces the SU(2) Skyrme model to that of a non-relativistic particle constrained over a sphere, a well known second-class problem\cite{sphere}\cite{NW3}.

The quantization of nonlinear constrained systems is a serious physical question that has been intensively studied over some decades by many authors\cite{many,Vers,Toda,Jan}. However, some kind of problems remains. For example, in the light of Dirac Hamiltonian formalism\cite{PD}, these models have field dependent brackets identified as quantum commutators. As established by the quantum mechanics, the quantum operators must be symmetrized adopting an ordering scheme. Since there are different acceptable prescriptions to construct a Hermitian operator, some of them may lead to different physical values, characterizing an operator ordering ambiguity. 

Recently, an alternative approach, based on the reformulation of a nonlinear model as a gauge invariant theory\cite{WN,BGB,BN,BW}, has been explored and some success has been achieved. In these papers, Wess-Zumino(WZ) variables were introduced on the theory, as suggest by Faddeev\cite{FS}, following different constraint conversion methods\cite{BT,IJMP}. 

In pioneer papers, two of us developed the reduced SU(2) Skyrme model as a gauge invariant theory using the BFFT formalism\cite{JW1,JW2}. These works inspired many authors\cite{HKP,NW,NW1} to investigate the gauge invariant version for the Skyrme model using different procedures. In these gauge-invariant formalisms, based on the Dirac's framework, the second-class constraints were converted into first-class ones with the introduction of the WZ variables. This process is affected by an ambiguity problem as shown in Ref.\cite{BN}. To overcome this kind of problem, we propose to use gauge-invariant formalisms which eliminate this arbitrariety. For example, the gauge unfixing Hamiltonian formalism\cite{MR,VT}. This formalism considers half (in the case of bosonic system) of total second-class constraints as gauge fixing terms while the remaining ones form a subset that satisfies a first-class algebra. However, this scheme is restrained to treat systems with even numbers of second-class constraints. In views of this, it is imperative to propose a new approach to carry out the gauge-invariant reformulation, namely, the symplectic gauge-invariant formalism. It is one of the main goals of this paper.

To prove that the symplectic gauge-invariant formalism
does not change the physical contents originally present on the second-class reduced SU(2) Skyrme model, the energy spectrum will be explicitly computed.
The result shows that this model may be described, in
the same phase-space coordinates, by both gauge invariant and noninvariant descriptions.

To become this paper self-consistent, it was organized as follow. In Section 2, we shall review the semi-classical expansion of the Skyrme's collective rotational mode. Reduction to a nonlinear quantum mechanical model depending explicitly on the time-dependent collective variables satisfying a spherical constraint is performed. In  section 3, the symplectic gauge-invariant formalism will be systematized, emphasizing the main steps and advantages. In section 4, we shall the hidden symmetry for the reduced SU(2) Skyrme model. To this end, this model will be reformulated as a gauge theory via symplectic gauge-invariant method and the infinitesimal gauge transformation will be computed. In Section 5, the gauge invariant system will be quantized employing the Dirac's first-class procedure, and the energy spectrum will be computed.  In Appendix, an alternative approach based on the gauge unfixing Hamiltonian method\cite{MR,VT} is shown to lead to canonically equivalent results. The last Section is reserved to discuss the physical meaning of our findings together with our final comments and conclusions.

\section{The reduced SU(2) Skyrme model}

The Skyrme model describes baryons and their interactions through soliton solution of the non-linear sigma model-type Lagrangian given by

\begin{eqnarray}
\label{Sky}
L = \int \, d^3x \[ -{F_\pi^2\over 16} Tr\, (\partial_\mu U \partial^u U^+) + {1 \over 32 e^2 } Tr [ U^+\partial_\mu U,
U^+\partial_\nu U ] ^2 \],
\end{eqnarray}

\noindent where $F_\pi$ is the pion decay constant, $e$ is a dimensionless parameter and $U$ is a SU(2) matrix. The collective semi-classical expansion\cite{Adkins}
is performed just substituting $U(x^\mu)$ by $U(x^\mu)=A(t)U(r)A^+(t)$ in (\ref{Sky}), where $A$ is a SU(2) matrix, we obtain

\begin{equation}
\label{Lag}
L = - M + \lambda Tr [ \partial_0 A\partial_0 A^{-1} ],
\end{equation}
where 

\begin{equation}
\label{mass}
M={F_\pi\over e} I_1
\end{equation}
and
\begin{equation}
\label{inertia}
\lambda={1\over e^3 F_\pi} I_2
\end{equation}
are the soliton mass and the moment of inertia respectively, and $I_1,I_2$ are adimensional values
depending on the classical solution of the model.
{\it A} is a SU(2) matrix which can be written as $A=a_0 +i a\cdot \tau$, where $\tau_i$ are the Pauli matrices, and satisfies the constraint relation

\begin{equation}
\label{pri}
T_1 = a_i a_i - 1 \approx 0, \,\,\,\, i=0,1,2,3.
\end{equation}

\noindent Then, the Lagrangian (\ref{Lag}) can be read as a function of the $a_i$ as

\begin{equation}
\label{cca}
L = -M + 2\lambda \dot{a}_i\dot{a}_i.
\end{equation}

\noindent Calculating the canonical momenta

\begin{equation}
\label{cm}
\pi_i = {\partial L \over \partial \dot{a}_i} = 4 \lambda \dot{a}_i,
\end{equation}
and using the Legendre transformation, the canonical Hamiltonian is computed as

\begin{eqnarray}
\label{chr}
H_c=\pi_i \dot a_i-L = M+2
 \lambda \dot a_i\dot a_i \nonumber \\
=M+{1\over 8 \lambda } \sum_{i=0}^3 \pi_i\pi_i.
\end{eqnarray}
A typical polynomial wave function,
${1\over N(l)}(a_1 + i a_2)^l = |polynomial \rangle\, ,$ is an 
eigenvector of the Hamiltonian (\ref{chr}). This wave function is also eigenvector of the spin and isospin operators, written in \cite{ANW} as $ J_k={1\over 2}
( a_0 \pi_k -a_k \pi_0 - \epsilon_{klm} a_l \pi_m )$  and 
$ I_k={1\over 2 } ( a_k \pi_0 -a_0 \pi_k- \epsilon_{klm} a_l\pi_m ).$

Constructing the total Hamiltonian and imposing that the constraint has no time evolution \cite{PD}, we get a new constraint

\begin{equation}
\label{T2}
T_2 = a_i\pi_i \approx 0 \,\,.
\end{equation}

\noindent We observe that no further constraints are generated via this iterative procedure because $T_1$ and $T_2$ are second-class constraints. The matrix elements of their Poisson brackets read

\begin{equation}
\label{Pa}
\Delta_{\alpha \beta} = \{T_\alpha,T_\beta\} = -2 \epsilon_{\alpha \beta}
a_i a_i, \,\, \alpha,\beta = 1,2
\end{equation}

\noindent where $\epsilon_{\alpha \beta}$ is the antisymmetric tensor normalized as $\epsilon_{12} = -\epsilon^{12} = -1$.
 
\section{Symplectic gauge-invariant formalism}

In the literature there are several schemes to reformulate noninvariant models as gauge theories. Recently, some constraint conversion formalisms, based on the Dirac's method\cite{PD}, were developed following the Faddeev's idea of phase-space extension with the introduction of auxiliary variables \cite{FS}. Among them, the BFFT\cite{BT} and the iterative\cite{IJMP} methods were powerful enough to be successfully applied to a great number of important physical models. Although these techniques share the same conceptual basis \cite{FS} and follow the Dirac's framework\cite{PD}, these constraint conversion methods were implemented following different directions. Historically, both BFFT and the iterative methods were applied to deal with linear systems such as chiral gauge theories\cite{IJMP,many3} in order to eliminate the gauge anomaly that hampers the quantization process. In spite of the great success achieved by these methods, they have an ambiguity problem\cite{BN}. This problem naturally arise when the second-class constraints is converted into first-class ones with the introduction of WZ variables. Due to this, the constraint conversion process may become a hard task\cite{BN}. In this section, we reformulate noninvariant systems as gauge theories using a new technique which is not affected by this ambiguity problem. This technique follows the Faddeev's suggestion\cite{FS} and is set up on a contemporary framework to handle noninvariant model,
namely, the symplectic formalism\cite{FJ,BC}. 

In order to systematize the symplectic gauge-invariant formalism, we consider a general noninvariant mechanical model whose dynamics is governed by a Lagrangian ${\cal L}(a_i,\dot a_i,t)$(with $i=1,2,\dots,N$), where $a_i$ and $\dot a_i$ are the space and velocities variables respectively. Notice that this model does not lead to lost generality or physical content. Following the symplectic method the Lagrangian is written in its first-order form as
 
\begin{equation}
\label{2000}
{\cal L}^{(0)} = A^{(0)}_\alpha\dot\xi^{(0)}_\alpha - V^{(0)},
\end{equation}
where $\xi^{(0)}_\alpha(a_i,p_i)$(with $\alpha=1,2,\dots,2N$) are the symplectic variables, $A^{(0)}_\alpha$ are the one-form canonical momenta, $(0)$ indicates that it is the zeroth-iterative Lagrangian and $V^{(0)}$ is the symplectic potential. After, the symplectic tensor, defined as

\begin{eqnarray}
\label{2010}
f^{(0)}_{\alpha\beta} = {\partial A^{(0)}_\beta\over \partial \xi^{(0)}_\alpha}
-{\partial A^{(0)}_\alpha\over \partial \xi^{(0)}_\beta},
\end{eqnarray}
is computed. Since this symplectic matrix is singular, it has a zero-mode $(\nu^{(0)})$ that generates a new constraint when contracted with the gradient of potential, namely,

\begin{equation}
\label{2020}
\Omega^{(0)} = \nu^{(0)}_\alpha\frac{\partial V^{(0)}}{\partial\xi^{(0)}_\alpha}.
\end{equation}
Through a Lagrange multiplier $\eta$, this constraint is introduced into the zeroth-iterative Lagrangian (\ref{2000}), generating the next one,

\begin{eqnarray}
\label{2030}
{\cal L}^{(1)} &=& A^{(0)}_\alpha\dot\xi^{(0)}_\alpha - V^{(0)}+ \dot\eta\Omega^{(0)},\nonumber\\
&=& A^{(1)}_\alpha\dot\xi^{(1)}_\alpha - V^{(1)},
\end{eqnarray}
where

\begin{eqnarray}
\label{2040}
V^{(1)}&=&V^{(0)}|_{\Omega^{(0)}= 0},\nonumber\\
\xi^{(1)}_\alpha &=& (\xi^{(0)}_\alpha,\eta),\\
A^{(1)}_\alpha &=& A^{(0)}_\alpha + \eta\frac{\partial\Omega^{(0)}}{\partial\xi^{(0)}_\alpha}.\nonumber
\end{eqnarray}
The first-iterative symplectic tensor is computed as

\begin{eqnarray}
\label{2050}
f^{(1)}_{\alpha\beta} = {\partial A^{(1)}_\beta\over \partial \xi^{(1)}_\alpha}
-{\partial A^{(1)}_\alpha\over \partial \xi^{(1)}_\beta}.
\end{eqnarray}
Since this tensor is nonsingular, the iterative process stops and the Dirac's brackets among the phase-space variables are obtained from the inverse matrix $(f^{(1)}_{\alpha\beta})^{-1}$. On the contrary, the tensor is singular and a new constraint arises, indicating that the iterative process goes on.

After this brief review, the symplectic gauge-invariant formalism will be systematized. It starts with the introduction of an extra term dependent on the original and WZ variable, $G(a_i,p_i,\theta)$, into the first-order Lagrangian. This extra term, expanded as

\begin{equation}
\label{2060}
G(a_i,p_i,\theta)=\sum_{n=0}^\infty{\cal G}^{(n)}(a_i,p_i,\theta),
\end{equation}
where ${\cal G}^{(n)}(a_i,p_i,\theta)$ is a term of order n in $\theta$, satisfies the boundary condition

\begin{eqnarray}
\label{2070}
G(a_i,p_i,\theta=0) = {\cal G}^{(n=0)}(a_i,p_i,\theta=0)=0.
\end{eqnarray}
The symplectic variables were extended to also contain the WZ variable $\tilde\xi^{(1)}_{\tilde\alpha} = (\xi^{(0)}_\alpha,\eta,\theta)$ (with ${\tilde\alpha}=1,2,\dots,2N+2$) and the first-iterative symplectic potential becomes

\begin{equation}
\label{2075}
{\tilde V}_{(n)}^{(1)}(a_i,p_i,\theta) = V^{(1)}(a_i,p_i) - \sum_{n=0}^\infty{\cal G}^{(n)}(a_i,p_i,\theta).
\end{equation}
For $n=0$, we have

\begin{equation}
\label{2075a}
{\tilde V}_{(n=0)}^{(1)}(a_i,p_i,\theta) = V^{(1)}(a_i,p_i).
\end{equation}
Subsequently, we impose that the symplectic tensor ($f^{(1)}$) is a singular matrix with the corresponding zero-mode 

\begin{equation}
\label{2076}
\tilde\nu^{(1)}_{\tilde\alpha}=\pmatrix{\nu^{(1)}_\alpha & 1},
\end{equation}
as the generator of gauge symmetry. Due to this, the correction terms  ${\cal G}^{(n)}(a_i,p_i,\theta)$ in order of $\theta$ can be explicitly computed. Contracting the zero-mode$(\tilde\nu^{(1)}_{\tilde\alpha})$ with the gradient of potential ${\tilde V}_{(n)}^{(1)}(a_i,p_i,\eta,\theta)$ and imposing that no more constraint is generated, a general differential equation is obtained, reads as

\begin{eqnarray}
\label{2080}
\tilde\nu^{(1)}_{\tilde\alpha}\frac{\partial {\tilde V}_{(n)}^{(1)}(a_i,p_i,\theta)}{\partial{\tilde\xi}^{(1)}_{\tilde\alpha}}&=&0,\nonumber\\
\nu^{(1)}_\alpha\frac{\partial V^{(1)}(a_i,p_i)}{\partial\xi^{(1)}_\alpha} - \sum_{n=0}^\infty\frac{\partial{\cal G}^{(n)}(a_i,p_i,\theta)}{\partial\theta}&=&0,
\end{eqnarray}
that allows us to compute all correction terms in order of $\theta$. For linear correction term, we have

\begin{equation}
\label{2090}
\nu^{(1)}_\alpha\frac{\partial V_{(n=0)}^{(1)}(a_i,p_i)}{\partial\xi^{(1)}_\alpha} - \frac{\partial{\cal
 G}^{(n=1)}(a_i,p_i,\theta)}{\partial\theta} = 0.
\end{equation}
For quadratic correction term, we get

\begin{equation}
\label{2095}
{\tilde\nu}^{(1)}_{\tilde\alpha}\frac{\partial V_{(n=1)}^{(1)}(a_i,p_i,\theta)}{\partial{\tilde\xi}^{(1)}_{\tilde\alpha}} - \frac{\partial{\cal
 G}^{(n=2)}(a_i,p_i,\theta)}{\partial\theta} = 0.
\end{equation}
From these equations, a recursive equation for $n\geq 1$ is proposed as

\begin{equation}
\label{2100}
{\tilde\nu}^{(1)}_{\tilde\alpha}\frac{\partial V_{(n-1)}^{(1)}(a_i,p_i,\theta)}{\partial{\tilde\xi}^{(1)}_{\tilde\alpha}} - \frac{\partial{\cal
 G}^{(n)}(a_i,p_i,\theta)}{\partial\theta} = 0,
\end{equation}
that allows us to compute each correction term in order of $\theta$. This iterative process is successively repeated until the equation (\ref{2080}) becomes identically null, consequently, the extra term $G(a_i,p_i,\theta)$ is obtained explicitly. Then, the gauge invariant Hamiltonian, identified as being the symplectic potential, is obtained as

\begin{equation}
\label{2110}
{\tilde{\cal  H}}(a_i,p_i,\theta) = V^{(1)}_{(n)}(a_i,p_i,\theta) = V^{(1)}(a_i,p_i) + G(a_i,p_i,\theta),
\end{equation}
and the zero-mode ${\tilde\nu}^{(1)}_{\tilde\alpha}$ is identified as being the generator of an infinitesimal gauge transformation, given by
\begin{equation}
\label{2120}
\delta{\tilde\xi}_{\tilde\alpha} = \varepsilon{\tilde\nu}^{(1)}_{\tilde\alpha},
\end{equation}
where $\varepsilon$ is an infinitesimal time-dependent parameter. 

In the next section, we reformulate the SU(2) Skyrme model as a gauge theory that, recently, has been intensively studied in the literature from many points of view\cite{Jan,JW1,JW2,HKP,NW1}, using the symplectic gauge-invariant process.

\section{Embedding the $SU(2)$ Skyrme model} 

In this section, the hidden symmetry of the reduced $SU(2)$ Skyrme model will be disclosed enlarging the phase-space with the introduction of the Wess-Zumino variable via symplectic gauge-invariant formalism. To put this work in a correct perspective, we first apply the symplectic method to the original second-class model, that allows us to show the second-class nature of the model and also to obtain the usual Dirac's brackets. Later, we unveil the hidden gauge symmetry of the model.

In order to implement the symplectic method, the original second-order Lagrangian in the velocity, given in (\ref{cca}), is reduced into a first-order form, namely,
 
\begin{equation}
\label{formula11}
L^{(0)} = \pi_i\dot{a}_i - M - \frac{1}{8\lambda}\pi_i\pi_i + \eta (a_ia_i - 1),
\end{equation}
where the index ${(0)}$ indicates the zeroth-iterative Lagrangian, and the Lagrange multiplier $(\eta)$ enforces the spherical constraint (\ref{pri}) into the theory. After, the symplectic tensor, defined as

\begin{eqnarray}
\label{tensor}
f_{\alpha\beta} = {\partial A_\beta\over \partial \xi^\alpha}
-{\partial A_\alpha\over \partial \xi^\beta},
\end{eqnarray}
must be computed. The zeroth-iterative symplectic variables are $\xi_\alpha^{(0)}=(a_j,\pi_j,\eta)$ and the corresponding one-form canonical momenta are given by

\begin{eqnarray}
\label{formula13}
A_{a_i}^{(0)} &=& \pi_i, \nonumber \\
A_{\pi_i}^{(0)} = A_{\lambda}^{(0)} &=& 0.
\end{eqnarray}
Then, the zeroth-iterative symplectic tensor is 

\begin{equation}
f^{(0)} = \left(
\begin{array}{ccc}
0           & -\delta_{ij} & 0 \\
\delta_{ij} &         0     & 0 \\
0           &         0     & 0
\end{array}
\right).
\end{equation}
This matrix is obviously singular, thus, it has a zero-mode

\begin{equation}
v^{(0)} = \left(
\begin{array}{ccc}
{\bf 0} \\
{\bf 0} \\
1
\end{array}
\right),
\end{equation}
that generates the following constraint

\begin{eqnarray}
\label{formula15}
\Omega_1 &=& v_\alpha^{(0)} \frac{\partial V^{(0)}}{\partial A_\alpha^{(0)}}, \nonumber \\
&=& a_ia_i - 1,
\end{eqnarray}
where the zeroth-iterative potential $V^{(0)}$ is

\begin{equation}
\label{15a}
V^{(0)} = M + \frac{1}{8\lambda}\pi_i\pi_i - \eta (a_ia_i - 1).
\end{equation}
Bringing back the constraint $\Omega_1$ into the canonical sector of the first-order Lagrangian by means of a Lagrange multiplier $\rho$, we get the first-iterative Lagrangian $L^{(1)}$, namely,

\begin{equation}
\label{formula16}
L^{(1)} = \pi_i\dot{a}_i + (a_ia_i - 1)\dot{\rho} - M - \frac{1}{8\lambda}\pi_i\pi_i,
\end{equation}
where $\eta\rightarrow \dot\rho$. Therefore, the symplectic variables become $\xi_\alpha^{(1)}=(a_j,\pi_j,\rho)$ with the following one-form canonical momenta

\begin{eqnarray}
\label{formula18}
A_{a_i}^{(1)} &=& \pi_i, \nonumber \\
A_{\pi_i}^{(1)} &=& 0,  \\
A_{\rho}^{(1)} &=& a_ia_i - 1.\nonumber
\end{eqnarray}
The corresponding matrix $f^{(1)}$ is 

\begin{equation}
f^{(1)}=\left(
\begin{array}{ccc}
0           & -\delta_{ij} & 2a_i \\
\delta_{ij} &         0     & 0 \\
-2a_i           &         0     & 0
\end{array}
\right),
\end{equation}
that is singular. The corresponding zero-mode is

\begin{equation}
v^{(1)}=\left(
\begin{array}{ccc}
{\bf 0} \\
a_i \\
1/2
\end{array}
\right),
\end{equation}
that generates the following constraint

\begin{eqnarray}
\label{formula20}
\Omega_2 &=& v_\alpha^{(1)} \frac{\partial V^{(1)}}
{\partial A_\alpha^{(1)}} \nonumber \\ \nonumber \\
&=& a_i\pi_i \approx 0,
\end{eqnarray}
where

\begin{equation}
\label{20a}
V^{(1)} = + M + \frac{1}{8\lambda}\pi_i\pi_i.
\end{equation}

The twice-iterated Lagrangian, obtained after including the constraint (\ref{formula20}) into the Lagrangian (\ref{formula16}) through a Lagrange multiplier $\zeta$, reads

\begin{equation}
\label{formula21}
L^{(2)} = \pi_i \dot{a}_i + (a_ia_i - 1)\dot{\rho}
+ a_i \pi_i \dot{\zeta} - V^{(2)},
\end{equation}
with $V^{(2)}$ = $V^{(1)}$. The enlarged symplectic variables are $\xi_\alpha^{(2)}=(a_j,\pi_j,\rho,\zeta)$. The new one-form canonical momenta are

\begin{eqnarray}
\label{formula22}
A_{a_i}^{(2)} &=& \pi_i, \nonumber \\
A_{\pi_i}^{(2)} &=& 0, \nonumber \\
A_{\rho}^{(2)} &=& a_ia_i - 1,\nonumber \\
A_{\zeta}^{(2)} &=& a_i \pi_i, \nonumber
\end{eqnarray}
and the corresponding matrix $f^{(2)}$ is

\begin{equation}
f^{(2)}=\left(
\begin{array}{cccc}
0           & -\delta_{ij}  &  2a_i   &   \pi_i \\
\delta_{ij} &         0     &   0    &    a_i \\
-2a_i         &         0     &   0    &     0   \\
-\pi_i       &        -a_i    &   0    &     0  
\end{array}
\right),
\end{equation}
that is a nonsingular matrix. The inverse of $f^{(2)}$ gives the usual Dirac brackets among the physical variables obtained in a straightforward calculation. This means that the $SU(2)$ Skyrme model is not a gauge invariant theory.

At this stage we are ready to implement our proposal. In order to disclose the hidden symmetry present on the reduced $SU(2)$ Skyrme model via symplectic gauge-invariant formalism, the original phase-space will be extended with the introduction of an extra function $G$ depending on the original phase-space variables and the WZ variable $\theta$, defined as

\begin{equation}
\label{23a}
G(a_i,\pi_i,\theta) = \sum_{n=0}^\infty {\cal G}^{(n)},
\end{equation}
that satisfies the boundary condition

\begin{equation}
\label{23b}
G(a_i,\pi_i,\theta=0) = {\cal G}^{(0)} = 0.
\end{equation}
Introducing the new term $G$ into the Lagrangian (\ref{formula16}), we have

\begin{equation}
\label{formula23}
\tilde L^{(1)} = \pi_i\dot{a}_i + ( a_i a_i -1)\dot{\rho} - M - \frac{1}{8\lambda}\pi_i\pi_i + G(a_i,\pi_i,\theta).
\end{equation}
The enlarged symplectic variables are $\tilde \xi_\alpha^{(1)}=(a_j,\pi_j,\rho,\theta)$
with the following one-form canonical momenta

\begin{eqnarray}
\label{formula25}
\tilde A_{a_i}^{(1)} &=& \pi_i, \nonumber \\
\tilde A_{\pi_i}^{(1)} &=& 0, \nonumber \\
\tilde A_{\rho}^{(1)} &=& a_ia_i - 1, \nonumber \\
\tilde A_{\theta}^{(1)} &=& 0.
\end{eqnarray}
Then, we compute the matrix $\tilde f^{(1)}$ as

\begin{equation}
\label{25a}
\tilde f^{(1)} = \left(
\begin{array}{cccc}
0           & -\delta_{ij}  &  2a_i   &   0 \\
\delta_{ij} &         0     &   0    &   0 \\
-2a_i         &         0     &   0    &     0   \\
0       &        0    &   0    &     0  
\end{array}
\right),
\end{equation}
that is obviously singular. Consequently, it has the following zero-mode

\begin{equation}
\label{25b}
v^{(1)}=\left(
\begin{array}{ccc}
{\bf 0} \\
a_i \\
1/2 \\
1
\end{array}
\right).
\end{equation}
Imposing that no more constraint is generated by this zero-mode $(v^{(1)})$, the first-order correction term in $\theta$, ${\cal G}^{(1)}$, is determined after an integration process, namely,

\begin{equation}
\label{formula26}
{\cal G}^{(1)}(a_i,\pi_i,\theta) = \frac{1}{4\lambda}(a_i\pi_i)\theta.
\end{equation}
Bringing back this expression into the Eq.(\ref{formula23}), the new Lagrangian is obtained as

\begin{equation}
\label{formula27}
\tilde L^{(1)} = \pi_i\dot{a}_i + ( a_ia_i - 1)\dot{\rho} - M - \frac{1}{8\lambda}\pi_i\pi_i + \frac{1}{4\lambda}(a_i\pi_i)\theta,
\end{equation}
that is not yet a gauge invariant Lagrangian because the zero-mode $v^{(1)}$ still generates a new constraint, reads as

\begin{equation}
\label{27a}
v^{(1)}_\alpha\frac{\partial V^{(1)}}{\partial\xi_\alpha} = + \frac {1}{4\lambda} a_i^2\theta,
\end{equation}
It indicates that is necessary to obtain the remaining correction terms ${\cal G}^{(n)}$ in order of $\theta$. It is achieved just imposing that the zero-mode does not generate more constraint. It allows us to determine the second-order correction term ${\cal G}^{(2)}$, given by

\begin{eqnarray}
\label{27b}
v^{(1)}_\alpha\frac{\partial V^{(1)}}{\partial\xi_\alpha} =  \frac {1}{4\lambda} a_i^2\theta + \frac{\partial{\cal
 G}^{(2)}}{\partial\theta} &=& 0,\nonumber\\
{\cal G}^{(2)} &=& - \frac {1}{8\lambda} a_i^2\theta.
\end{eqnarray}
Bringing this result into the first-order Lagrangian (\ref{formula27}), we obtain

\begin{equation}
\label{formula28a}
\tilde L^{(1)} = \pi_i\dot{a}_i + ( a_ia_i - 1)\dot{\rho} - M - \frac{1}{8\lambda}\pi_i\pi_i + \frac{1}{4\lambda}(a_i\pi_i)\theta - \frac
 {1}{8\lambda} a_i^2\theta.
\end{equation}
The zero-mode $v^{(1)}$ does not produce a new constraint, consequently, the model has a symmetry and it is the generator of an infinitesimal gauge transformation. Due to this, all correction terms ${\cal G}^{(n)}$ with $n\geq 3$ are nulls.

At this moment, we are interested to recover the invariant second-order Lagrangian from its first-order  form given in Eq.(\ref{formula28a}). To this end, the canonical momenta must be eliminated from the Lagrangian (\ref{formula28a}). From the equation of motion for $\pi_i$, the canonical momenta is computed as

\begin{equation}
\label{28b1}
\pi_i = 4\lambda\dot a_i + a_i\theta.
\end{equation}
Inserting this result into the first-order Lagrangian, given by
\begin{equation}
\label{formula27a}
\tilde L^{(0)} = \pi_i\dot{a}_i + ( a_ia_i - 1){\eta} - M - \frac{1}{8\lambda}\pi_i\pi_i + \frac{1}{4\lambda}(a_i\pi_i)\theta - \frac
 {1}{8\lambda}a_i^2\theta^2,
\end{equation}
the second-order Lagrangian is obtained as

\begin{equation}
\label{26b}
\tilde L = - M + 2\lambda\dot{a}_i^2 + (a_i\dot a_i)\theta + (a_ia_i - 1)\eta,
\end{equation}
with the corresponding gauge invariant Hamiltonian

\begin{equation}
\label{formula28}
\tilde H = M + \frac{1}{8\lambda}\pi_i \pi_i - \frac{1}{4\lambda}(a_i\pi_i)\theta + \frac {1}{8\lambda}a_i^2\theta^2 - \eta (a_ia_i -1).
\end{equation}
By construction, both Lagrangian (\ref{26b}) and Hamiltonian (\ref{formula28}) are gauge invariant. 

To become this work self-consistent the infinitesimal gauge transformation will be determined using the symplectic method. To this end, we start with the first-order Lagrangian (\ref{formula28a}) in terms of the symplectic variables $\tilde \xi_\alpha^{(1)}=(a_j,\pi_j,\rho,\theta)$, that generates the singular symplectic matrix (\ref{25a}) with the zero-mode (\ref{25b}). This zero-mode is identified as being the generator of the infinitesimal gauge transformation $\delta\tilde \xi_\alpha^{(1)}=\varepsilon v^{(1)}$, given by

\begin{eqnarray}
\label{28a}
\delta a_i &=& 0,\nonumber\\
\delta \pi_i &=& \varepsilon a_i ,\nonumber\\
\delta \eta &=& \frac 12\dot\varepsilon,\,\,(\eta\rightarrow \dot \rho)\\
\delta \theta &=& \varepsilon.\nonumber\\
\end{eqnarray}
Note that both Hamiltonian and Lagrangian are invariant under this transformation. Similar results were also obtained in the literature using different methods based on the Dirac's constraint idea\cite{WN,BGB,JW1,JW2,HKP,NW}. However, these methods are affected by some ambiguities problems that naturally arise when it is necessary to obtain the second-class constraints and then determine how they will be converted into first-class ones. It occurs when the phase-space is extended with the introduction of the WZ variables. In our procedure, this kind of problem does not arise, consequently, the arbitrariety disappears. This completes  one of the main goal of this paper.

Henceforth we are interested to disclose the hidden symmetry of the reduced $SU(2)$ Skyrme model and obtain both Hamiltonian and  Lagrangian in terms of the original coordinates $(a_i,\pi_i)$. To this end, we will obtain the set of constraints of the invariant model described by the Lagrangian (\ref{26b}) and Hamiltonian (\ref{formula28}). Indeed, the model has two constraint chains, namely,

\begin{eqnarray}
\label{28b}
\phi_1 &=& \pi_\eta,\nonumber\\
\phi_2 &=& a_ia_i - 1,
\end{eqnarray}
and

\begin{eqnarray}
\label{28c}
\varphi_1 &=& \pi_\theta,\nonumber\\
\varphi_2 &=& a_i\pi_i - a_i^2\theta,
\end{eqnarray}
where $\pi_\theta$ is the canonical momentum conjugated to the WZ variable $\theta$. The Dirac matrix is singular, however, there are nonvanishing Poisson brackets among some constraints, indicating that there are both second-class and first-class constraints. It is solved splitting up the second-class constraints from the first-class ones through the constraints combination. The set of first-class constraints is

\begin{eqnarray}
\label{28d}
\chi_1 &=& \pi_\eta,\nonumber\\
\chi_2 &=& a_ia_i - 1 - 2\pi_\theta,
\end{eqnarray}
while the set of second-class constraints is

\begin{eqnarray}
\label{28e}
\chi_1 &=& \pi_\theta,\nonumber\\
\chi_1 &=& a_i\pi_i - a_i^2\theta.
\end{eqnarray}
Since the second-class constraints are assumed in a strong way, and using the Maskawa-Nakajima theorem\cite{NM}, the Dirac's brackets are worked out as 

\begin{eqnarray}
\label{28f}
\lbrace a_i, a_i\rbrace &=& 0,\nonumber\\
\lbrace a_i, p_i\rbrace &=& \delta_{ij},\\
\lbrace p_i, p_i\rbrace &=& 0,\nonumber
\end{eqnarray}
as well as the Hamiltonian,

\begin{eqnarray}
\label{28g}
\tilde H &=& M + \frac{1}{8\lambda}\pi_i \pi_i - \frac{1}{8\lambda}\frac{(a_i\pi_i)^2}{a_ia_i} - \eta (a_ia_i -1)\nonumber\\
&=& M+{1\over 8\lambda} \pi_iM_{ij}\pi_j- \eta (a_ia_i -1),
\end{eqnarray}
where

\be
\label{matrix1}
M_{ij} = \delta_{ij} - \frac {a_ia_j}{a_k^2},
\ee
is a singular matrix. We can show that $\tilde{H}$, Eq.(\ref{28g}), satisfies the first-class property

\begin{equation}
\label{Htilde1}
\{ T_1, \tilde{H} \} = 0.
\end{equation}
Due to this the first-class  constraint ($T_1$) is the generator of the gauge symmetry. The infinitesimal gauge transformation are computed as

\ba
\label{transf1}
\delta a_i &=& \varepsilon\lbrace a_i,T_1\rbrace = 0,\nonumber\\
\delta \pi_i &=& \varepsilon\lbrace\pi_i,T_1\rbrace=
\varepsilon a_i,\nonumber\\
\ea
where $\varepsilon$ is an infinitesimal time-dependent parameter. It is easy to verify that the Hamiltonian (\ref{28g}) is invariant under these transformations because $a_i$ are eigenvectors of the phase space metric ($M_{ij}$) with eigenvalue nulls.
It reproduces the result discussed in appendix using the gauge unfixing Hamiltonian formalism. 

\section{ The spectrum of the Hamiltonian}

In this section, we will derive the SU(2) Skyrmion energy levels. Normally, these results were employed to obtain the baryons static properties\cite{Adkins,ANW}. In this first-class theory the quantization is performed following the Dirac's first-class prescription\cite{PD} just imposing that the physical wave functions are annihilated by the first-class operator constraint, reads as

\begin{eqnarray}
\label{qope}
\xi |\psi \rangle_{phys} = 0.
\end{eqnarray}
The physical states that satisfy (\ref{qope}) is

\begin{eqnarray}
\label{physical}
| \psi \rangle_{phys} = {1\over V } \, 
\delta(a_i a_i-1)\,|polynomial\rangle.
\end{eqnarray}

\noindent where {\it V } is the normalization factor and $|polynomial \rangle ={1\over N(l)} (a_1+ i a_2)^l \,$. The corresponding quantum Hamiltonian is

\begin{eqnarray}
\label{echs1}
\tilde{H}= M+{1\over 8\lambda} \pi_iM_{ij}\pi_j- \eta (a_ia_i -1).
\end{eqnarray}
Thus, in order to obtain the spectrum of the theory, we take the scalar product, 
$_{phys}\langle\psi| \tilde{H} | \psi \rangle_{phys}\,$, that is the mean value of the first- class Hamiltonian. We begin calculating the scalar product, given by

\begin{eqnarray}
\label{mes1}
_{phys}\langle\psi| \tilde{H} | \psi \rangle_{phys}=\nonumber \\
\langle polynomial |\,\,  {1\over V^2}  \int da_i\,\,
\delta(a_i a_i - 1)\,
\tilde{H}\,
\delta(a_i a_i - 1)\,\,
| polynomial \rangle .
\end{eqnarray}

\noindent Integrating over $a_i$, we obtain

\begin{eqnarray}
\label{mes2}
_{phys}\langle\psi| \tilde{H} | \psi \rangle_{phys}=\nonumber \\
\langle polynomial | M + {1\over 8\lambda}\[ \pi_i \pi_i 
- (a_i\pi_i)^2 \] | polynomial \rangle.
\end{eqnarray}

\noindent Here we would like to comment that the regularization of delta function squared 
$\delta(a_i a_i - 1 )^2$ is performed using the delta relation,  $(2\pi)^2\delta(0)=\lim_{k\rightarrow 0}\int d^2x \,e^{ik\cdot x} =\int d^2x= V.$ Then, we use the parameter V as the normalization factor. The Hamiltonian operator inside the kets, Eq.(\ref{mes2}), can be rewritten as

\begin{eqnarray}
\label{mes3}
_{phys}\langle\psi| \tilde{H} | \psi \rangle_{phys}=\nonumber \\
\langle polynomial | M + {1\over 8\lambda} \[ p_k\cdot p_k \] | polynomial \rangle,
\end{eqnarray}

\noindent where $p_k=\pi_k - a_k (a_j\pi_j)$. The operator $\pi_k$ describe a free particle and their representations on the collective coordinates $a_k$ are

\begin{eqnarray}
\label{collec}
\pi_k= -i {\partial\over \partial a_k}.
\end{eqnarray}

\noindent The algebraic expression of  $p_k$ lead to ordering problems in the first-class Hamiltonian operator $\tilde{H}$.  We adopt the  well known Weyl ordering prescription\cite{Weyl} to symmetrized the $p_k$ expression, and consequently $\tilde{H}$. We count all possible randomly order of $\pi_i$ and $a_k$. Then, the symmetrized expressions for $p_k$ are

\begin{eqnarray}
\label {psym}
[p_k]_{sym} & = &{1\over 6i} (6\partial_k
-a_k a_i\partial_i - a_k\partial_i a_i
- a_i a_k\partial_i - a_i\partial_i a_k\nonumber\\ 
&-&\partial_i a_k a_i
-\partial_i a_i a_k)\nonumber\\ \nonumber \\
& = & {1\over i} \( \partial_k-a_k a_i \partial_i - {5\over 2} a_k \),
\end{eqnarray}

\noindent leading to the symmetrized first-class Hamiltonian operator

\begin{eqnarray}
\label{Hsym}
[\tilde{H}]_{sym} = M+{1\over 8\lambda}\[ -\partial_j\partial_j+{1\over2}\(
 OpOp+2Op+{5\over4} \)\],
\end{eqnarray}

\noindent where $Op$ is defined as $Op\equiv a_i\partial_i$. Putting the expression (\ref{Hsym}) in the mean value, (\ref{mes3}), we obtain the energy levels as

\begin{eqnarray}
\label{energy}
E_l=_{phys}\langle\psi| \tilde{H} | \psi \rangle_{phys}=M+{1\over 8\lambda} \[ l(l+2)+{5\over 4} \].
\end{eqnarray}

\noindent We would like to comment that the last expression, Eq.(\ref{energy}), matches with the result obtained in Ref.\cite{Jan}, where the SU(2) Skyrme model was quantized via second-class Dirac's method. It becomes an interesting point since this extra term play an important role in the energy Skyrmion spectrum\cite{HKP}. 
It can be shown just observing in Eq.(\ref{energy}) that the value of the
soliton mass ($M$), Eq.(\ref{mass}), and the inertia moment ($\lambda$), Eq.(\ref{inertia}), 
are determined using the nucleon (l=1) and the delta (l=2) masses as input parameters. Consequently, the values of $F_\pi$, $e$, and the remaining phenomelogical results can be predicted.
Then, it is clear that an extra term,  resulting from a second or first-class quantization scheme together with a symmetrization procedure, can modify the spectrum and, therefore, the physical values predicted by the Skyrme model.
In the context of the non-Abelian and Abelian BFFT formalisms(used by two of us in early papers\cite{JW1,JW2}) the extra constant term in the energy formula, Eq.(\ref{energy}), does not match with the obtained in the second-class formalism\cite{Jan}.

\section{Final Discussions}

In this paper, we propose a new gauge-invariant formalism that is no affected by an ambiguity problem related to the introduction of the WZ variables. This formalism was systematized and applied on the reduced SU(2) Skyrme model. The hidden symmetry living on the original phase-space was investigated which is an unexpected result for a second-class system. Afterward, this invariant model was quantized employing the Dirac's first-class procedure. Using the Weyl ordering prescription to symmetrize the operators, we obtained exactly the same energy spectrum when compared with the reduced second-class Skyrme model. It is an important feature that not occur when the BFFT method is used\cite{JW1,JW2,HKP}. We believe that the arbitrary algebra in the extended model, induced by the introduction of the Wess-Zumino variables, leads to the discrepancy between the first and the second-class Skyrmion energy spectrum. In view of this, different constraint conversion schemes introduce distinct modifications in the energy spectrum\cite{JW1,JW2,HKP} and, consequently, change the phenomenological results, as discussed in section V.

Our results prove that the SU(2) Skyrme model has also a gauge invariant description (on the original phase-space coordinates $(a_i, p_i)$) dynamically equivalent to the usual second-class treatment. It seems important since our scheme does not affect the baryon phenomenology initially predicted by the second-class model, in opposition to another gauge-invariant formalism\cite{JW1,JW2,HKP,NW,NW1}. Thus, the symplectic gauge-invariant  formalism leads to a more elegant and simplified first-class Hamiltonian structure than the Abelian and non-Abelian BFFT cases.

\section{ Acknowledgments}
This work is supported in part by FAPEMIG, Brazilian Research Council.  In particular, C. Neves would like
to  acknowledge the FAPEMIG grant no. CEX-00005/00.
 
\section{Appendix}

\subsection{ The gauge unfixing formalism for the reduced SU(2) Skyrme model}

The main idea of the gauge unfixing procedure is to consider half of the total second-class constraints as gauge fixing terms while the remaining ones are the gauge symmetry generators\cite{VT,HT}. Here, the gauge unfixing Hamiltonian formalism will be applied to the reduced SU(2) Skyrme model reviewed in the section 2. We start redefining the constraint $T_1=a_ia_i-1$ as

\begin{eqnarray}
\label{red}
\xi=C^{-1} T_1,
\end{eqnarray}

\noindent where $C$ is 

\begin{eqnarray}
\label{C}
C= \{T_1, T_2 \}= 2 a_i a_i=2.
\end{eqnarray} 

\noindent After, the total Hamiltonian is 
written as

\begin{eqnarray}
\label{Htotal}
H= M+{1\over 8\lambda} \pi_i\pi_i +\eta_1\xi+\eta_2 T_2,
\end{eqnarray}
where $\eta_1$ and $\eta_2$ are the Lagrange multipliers that enforces the constraints $\xi$ and $T_2$ into the Hamiltonian. Imposing that the constraints $\xi$ and $\psi$ are conserved on time, the Lagrange multipliers are obtained as

\begin{eqnarray}
\label{mult1}
\eta_1= {1\over 4\lambda} \pi_i\pi_i,\\
\label{mult2}
\eta_2= -{1\over 4\lambda} a_i\pi_i.
\end{eqnarray}

\noindent Substituting Eq.(\ref{mult1}) and Eq.(\ref{mult2}) in the total Hamiltonian given in Eq.(\ref{Htotal}), we get

\begin{eqnarray}
\label{Htotal2}
H= M+{1\over 8\lambda} \pi_i\pi_i - {1\over 4\lambda} 
(a_i\pi_i)^2.
\end{eqnarray}
Then, we are ready to derive the gauge invariant Hamiltonian using the formula\cite{VT}, given by

\begin{eqnarray}
\label{HF}
\tilde{H} = H - \psi \{ \xi, H \} + {1\over 2!} \psi^2 \{ \xi ,  \{ \xi, H \} \}
-  {1\over 3!} \psi^3 \{\xi, \{ \xi , \{ \xi, H \} \} + \dots .
\end{eqnarray}

\noindent The right side hand terms $\{\xi,H\}$ and $\{\xi,\{\xi,H\}\}$ are computed,

\begin{eqnarray}
\label{b1}
\{\xi,H\}= -{1\over 4\lambda} a_i\pi_i,\\
\label{b2}
\{\xi,\{\xi,H\}\}= -{1\over 4\lambda}.
\end{eqnarray}

\noindent  From Eq.(\ref{b2}) we note that the terms in (\ref{HF}), $\{\xi,\{\xi,\{\xi,H\}\}\}$, and the remainders higher orders are zero. Then, the invariant Hamiltonian reads

\begin{eqnarray}
\label{Fclass}
\tilde{H}&=& M+{1\over 8\lambda} \pi_i\pi_i - {1\over 8\lambda} 
(a_i\pi_i)^2,\nonumber\\
&=& M+{1\over 8\lambda} \pi_i{\bar M}_{ij}\pi_j,
\end{eqnarray}
where

\be
\label{matrix}
{\bar M}_{ij} = \delta_{ij} - a_ia_j,
\ee
is a singular matrix. We can show that $\tilde{H}$, Eq.(\ref{Fclass}), satisfies the first-class property

\begin{equation}
\label{Htilde}
\{ \xi, \tilde{H} \} = 0.
\end{equation}
Due to this the first-class  constraint ($\xi$) is the generator of the gauge symmetry. The infinitesimal gauge transformations are computed as

\ba
\label{transf}
\delta a_i &=& \varepsilon\lbrace a_i,\xi\rbrace = 0,\nonumber\\
\delta \pi_i &=& \varepsilon\lbrace\pi_i,\xi\rbrace=\varepsilon a_i,
\ea
where $\varepsilon$ is an infinitesimal time-dependent parameter. It is easy to verify that the Hamiltonian (\ref{Fclass}) is invariant under these transformations because $a_i$ are eigenvectors of the phase space metric (${\bar M}_{ij}$) with eigenvalues nulls.

To complete this section, we would like to remark that the algebraic expression for the first-class Hamiltonian, Eq.(\ref{Fclass}), is more
simple than obtained via the Abelian and non-Abelian BFFT formalism as shown by two of us in Ref.\cite{JW1,JW2}. In the context of Abelian formalism\cite{JW1}, the first-class Hamiltonian has a geometrical series form, while in the non-Abelian formalism\cite{JW2,HKP} the first-class Hamiltonian has
a finite numbers of terms, but this algebraic formula is large.

\end{document}